\documentclass[twocolumn,resetfootnote]{aastex701}

\begin{document}

\title{Intertwined birth and death: a Herbig–Haro outflow resolves the distance to Vela Junior}

\author[orcid=0000-0001-7068-9702,sname='Suherli']{Janette Suherli}
\affiliation{University of Manitoba, Department of Physics and Astronomy}
\email[show]{suherlij@myumanitoba.ca}  

\author[orcid=0000-0002-5044-2988]{Ivo R. Seitenzahl} 
\affiliation{Australian National University, Mathematical Sciences Institute}
\affiliation{Australian National University, Research School of Astronomy and Astrophysics}
\email{ivoseitenzahl@gmail.com}

\author[orcid=0000-0001-6189-7665]{Samar Safi-Harb}
\affiliation{University of Manitoba, Department of Physics and Astronomy}
\email{Samar.Safi-Harb@umanitoba.ca}

\author[orcid=0000-0002-9665-2788]{Frédéric P. A. Vogt}
\affiliation{Federal Office of Meteorology and Climatology MeteoSwiss}
\email{frederic.vogt@alumni.anu.edu.au}

\author[orcid=0000-0002-6089-6836]{Wynn C. G. Ho}
\affiliation{Department of Physics and Astronomy, Haverford College, 370 Lancaster Avenue, Haverford, PA 19041, USA}
\email{who@haverford.edu}

\author[orcid=0000-0002-9886-0839]{Parviz Ghavamian}
\affiliation{Towson University, Department of Physics, Astronomy and Geosciences}
\email{pghavamian@towson.edu}

\author[orcid=0000-0003-1449-7284]{Chuan-Jui Li}
\affiliation{National Chengchi University, Graduate Institute of Applied Physics}
\email{cjli@g.nccu.edu.tw}

\author[orcid=0000-0002-4794-6835]{Ashley J. Ruiter}
\affiliation{Australian National University, Mathematical Sciences Institute}
\email{ashley.ruiter@gmail.com}

\author[orcid=0000-0002-2036-2426]{Roland M. Crocker}
\affiliation{Australian National University, Research School of Astronomy and Astrophysics}
\email{roland.crocker@anu.edu.au}

\author[orcid=0000-0002-5021-6737]{Arpita Roy}
\affiliation{Australian National University, Research School of Astronomy and Astrophysics}
\email{arpita.roy1016@gmail.com}

\author[orcid=0000-0002-6620-7421]{Ralph Sutherland}
\affiliation{Australian National University, Research School of Astronomy and Astrophysics}
\email{Ralph.Sutherland@anu.edu.au}


\begin{abstract}
The distance to the Vela Junior supernova remnant (RX J0852.0--4622 or G266.2--1.2) has long remained uncertain, limiting our understanding of its physical properties. Using VLT/MUSE integral field spectroscopy, we uncover chemical and kinematic connections between the nebula surrounding its Central Compact Object (CXOU J085201.4--461753) and the nearby Herbig-Haro outflow of Ve~7--27 (Wray~16--30), indicating a shared nitrogen-rich, Fe-peak-enhanced environment. This link ties stellar birth and death, with the young star Ve~7--27 embedded in material expelled by Vela Junior's massive progenitor, and the remnant's blast wave is expanding through the same medium. Adopting the \textit{Gaia}-based distance to Ve~7--27, we revise Vela Junior’s distance to $1.41\pm0.14$~kpc. At this distance, the remnant's physical radius is $23.3\pm2.3$~pc, and X-ray proper motions of the northwestern rim correspond to shock speeds of $(2.8\pm0.7)\times10^3$ to $(5.6\pm1.5)\times10^3$~km\,s$^{-1}$. These imply an age of $\sim$1.6--3.3~kyr and a very low ambient density, indicating that Vela Junior is expanding within a highly rarefied wind-blown cavity carved by a massive progenitor -- consistent with the non-detection of strong thermal X-ray emission. This distance update also resolves long-standing inconsistencies, with major implications for its energy budget, particle acceleration efficiency, and compact object evolution.

\end{abstract}



\section{Introduction} \label{sec:intro}

Massive stars and their deaths both seed and sculpt galaxies: they form in dense molecular clouds, enrich the interstellar medium (ISM) through their winds, and end their lives as supernovae that inject energy and heavy elements, drive fast shocks, and accelerate cosmic rays. Supernova remnants (SNRs) are therefore prime laboratories for studying these processes. However, interpretations are often limited by uncertain distance estimates, which set the fundamental scale for ages, energetics, and particle acceleration. 

Vela Junior (RX J0852.0--4622, G266.2--1.2, hereafter Vela~Jr) exemplifies this problem. Despite more than two decades of study, its distance has remained remarkably uncertain, with published estimates spanning over an order of magnitude (Figure~\ref{fig:distance_compilation}; see Appendix~\ref{app:distance_review}). This wide spread has led to mutually inconsistent inferences for its intrinsic properties. Early work suggested that Vela~Jr is an exceptionally nearby and young remnant, whereas later analyses based on X-ray absorption modeling, shock proper-motions, and the properties of its central compact object (CCO) favored substantially larger distances. All of these estimates, however, are inherently model-dependent and often probe only the northwestern rim.

\begin{figure*}[ht!]
\plotone{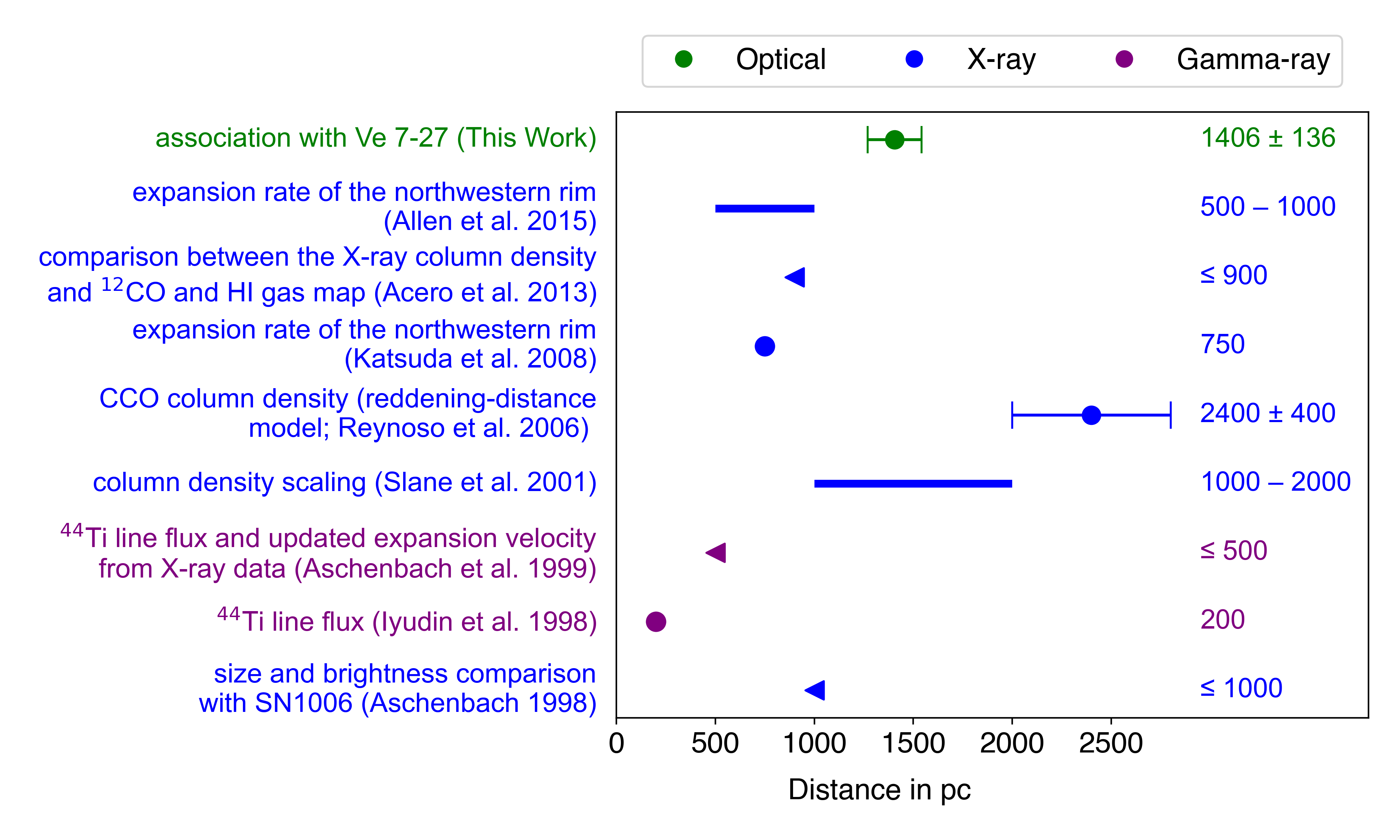}
\caption{Distance estimates to Vela~Jr from the literature (see Appendix~\ref{app:distance_review}) and this work, arranged in reverse chronological order. The y-axis lists the methods used, and the colors indicate the wavelength regime of the observations used in the estimate: green for optical, blue for X-ray, and purple for $\gamma$-ray.}
\label{fig:distance_compilation}
\end{figure*}

Discovered in \textit{ROSAT} hard X-ray data in 1998 \citep{Aschenbach1998}, Vela~Jr is one of the few synchrotron-dominated SNRs \citep{Slane2001, 2008ARA&A..46...89R, 2012AdSpR..49.1313F} and a prominent TeV $\gamma$-ray source \citep{Aharonian2005,Aharonian2007,HESS2018}. It has been proposed as a nearby cosmic-ray accelerator \citep{Fukui2017,Fukui2024} and a potential continuous gravitational-wave emitter \citep{2022PhRvD.105h2005A,Ming2024}. Pinning down its distance accurately is therefore essential not only for interpreting Vela~Jr's properties, but also for constraining models of stellar evolution, particle acceleration, nucleosynthesis, and feedback in the ISM. 

Vela~Jr hosts the CCO CXOU~J085201.4--461753 \citep{Slane2001, Pavlov2001}, the only CCO known with a detected optical nebula \citep{Pellizzoni2002, Mignani2007, Suherli2024}. CCOs, typified by the one in Cas~A \citep{Tananbaum1999}, form a rare class of young neutron stars that emit almost exclusively in the soft X-ray band and lack observable pulsar wind nebulae, radio pulsations, or $\gamma$-ray emission \citep{DeLuca2017}. The optical nebula surrounding the Vela~Jr CCO was initially identified as an H$\alpha$ nebula, possibly a bow shock or photoionized cloud \citep{Pellizzoni2002,Mignani2007}. MUSE observations revealed it to be 8\arcsec\! wide, dominated by [N\,{\sc ii}]$\lambda\lambda$6548,6583 emission \citep{Suherli2024}, with a faint, previously undetected arc of H$\alpha$ along its southern edge. Photoionization modeling shows the nebula is consistent with nitrogen-rich gas illuminated by a hot source consistent with the CCO thermal radiation. The extreme nitrogen ([N\,{\sc ii}]$\lambda\lambda$6548+6583/H$\alpha$$\sim$34) composition of the gas matches the expectations from the final wind phase of a massive nitrogen-rich Wolf-Rayet (WN) star \citep{Roy2020}. This fossil wind, now lit by the CCO, indicates that Vela~Jr formed from the collapse of a WN-type star that shed much of its outer layers before the supernova, enriching its surroundings with stellar material.

In this Letter, we present the Multi Unit Spectroscopic Explorer \citep[MUSE;][]{Bacon2010} observations of the central region of Vela~Jr (Program ID 0104.D-0092(B); P.I.: F.P.A. Vogt; see Appendix~\ref{app:muse_data_desc}), covering both the CCO nebula and the bright emission object Ve~7--27\footnote{Sometimes listed erroneously as Ve~2--27.} (Wray 16--30), located $\sim$22\arcsec\, from the CCO. The full MUSE mosaic provides a comprehensive view of the CCO nebula and Ve~7--27, including its bipolar outflow and its interaction with the surrounding nitrogen-enriched medium. In Section~\ref{sec:ve727isHH}, we show that Ve~7--27 is a Herbig-Haro object with pronounced nitrogen enrichment and in Section~\ref{sec:revisedDistance}, we demonstrate the chemical and kinematic associations between Ve~7--27 and the CCO nebula, which enables a revised, model-independent distance to Vela~Jr. Section~\ref{sec:discussion} discusses the implications of this new distance for the remnant's physical parameters and the broader evolutionary context of the system, and Section~\ref{sec:conclusions} summarizes our conclusions.

\section{Ve~7--27: A Herbig-Haro Outflow} \label{sec:ve727isHH}

Ve~7--27 is a bright emission-line source with a \textit{Gaia} parallax distance of 1406$\pm$136~pc \citep{Gaia2020}. Its classification has been debated: initially identified as a compact planetary nebula (PN) candidate \citep{Velghe1957}, later reclassified as a Be star \citep{Wray1966} with near-infrared excess \citep{Allen1975}, and subsequently reported to show low-ionization optical spectral features reminiscent of $\eta$ Carinae\footnote{$\eta$ Carinae is a Luminous Blue Variable (LBV) object.} \citep{Landaberry2001}. Such spectra indicate dense circumstellar material, but are not uniquely diagnostic of the underlying stellar type. Follow-up radio observations again favored the PN interpretation \citep{Reynoso2006}.

\begin{figure*}[ht!]
\plotone{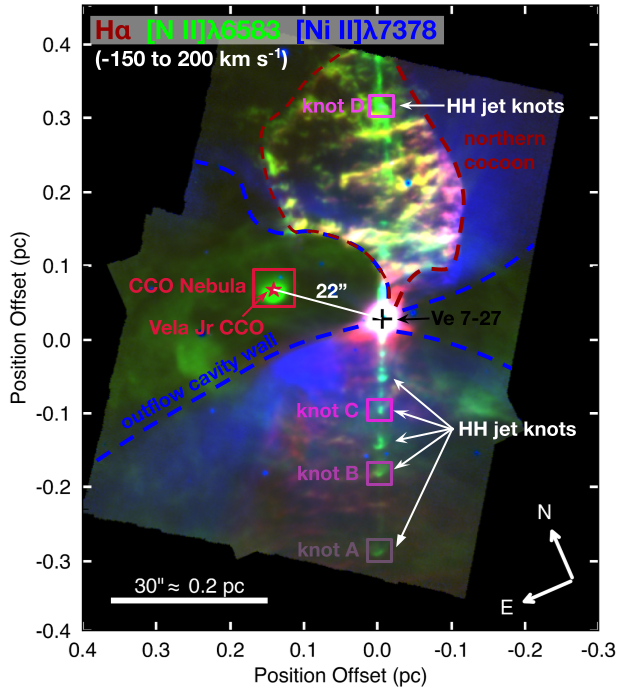}
\caption{Continuum-subtracted MUSE RGB composite of the central region of Vela~Jr in H$\alpha$ (R), [N\,{\sc ii}]$\lambda$6583 (G), and [Ni\,{\sc ii}]$\lambda$7378 (B), integrated from $-150$~km\,s$^{-1}$ to $+200$~km\,s$^{-1}$. The image is rotated by 23.5$^{\circ}$ counter-clockwise to place the jet axis vertically. Key structure of the Ve~7--27 HH system are annotated. }
\label{fig:muse_annotated}
\end{figure*}

MUSE provides the first spatially resolved view of Ve~7–27's optical structure, revealing features inconsistent with those of a typical PN. Figure~\ref{fig:muse_annotated} shows an RGB composite of H$\alpha$ (R), [N\,{\sc ii}]$\lambda$6583 (G), and [Ni\,{\sc ii}]$\lambda$7378 (B), integrated from $-150$~km\,s$^{-1}$ to $+200$~km\,s$^{-1}$ and rotated so that the jet axis aligns with the y-axis. Ve~7--27 displays a narrow bipolar jet composed of well-defined strings of shock-excited knots. Distinct bow shock structures are observed particularly along the southern jet, while the northern jet is enclosed by a conical cocoon of gas. The bipolar jet extends at least $\sim$47\arcsec\! on each side, corresponding to a projected physical length of 0.32~pc. This morphology is characteristic of Herbig-Haro (HH) objects, which trace highly collimated outflows driven by dense shock interactions in the immediate environment of young stellar objects (YSOs) \citep[for reviews, see e.g.][]{Reipurth2001, Bally2016}. 

\begin{figure*}[ht!]
\plotone{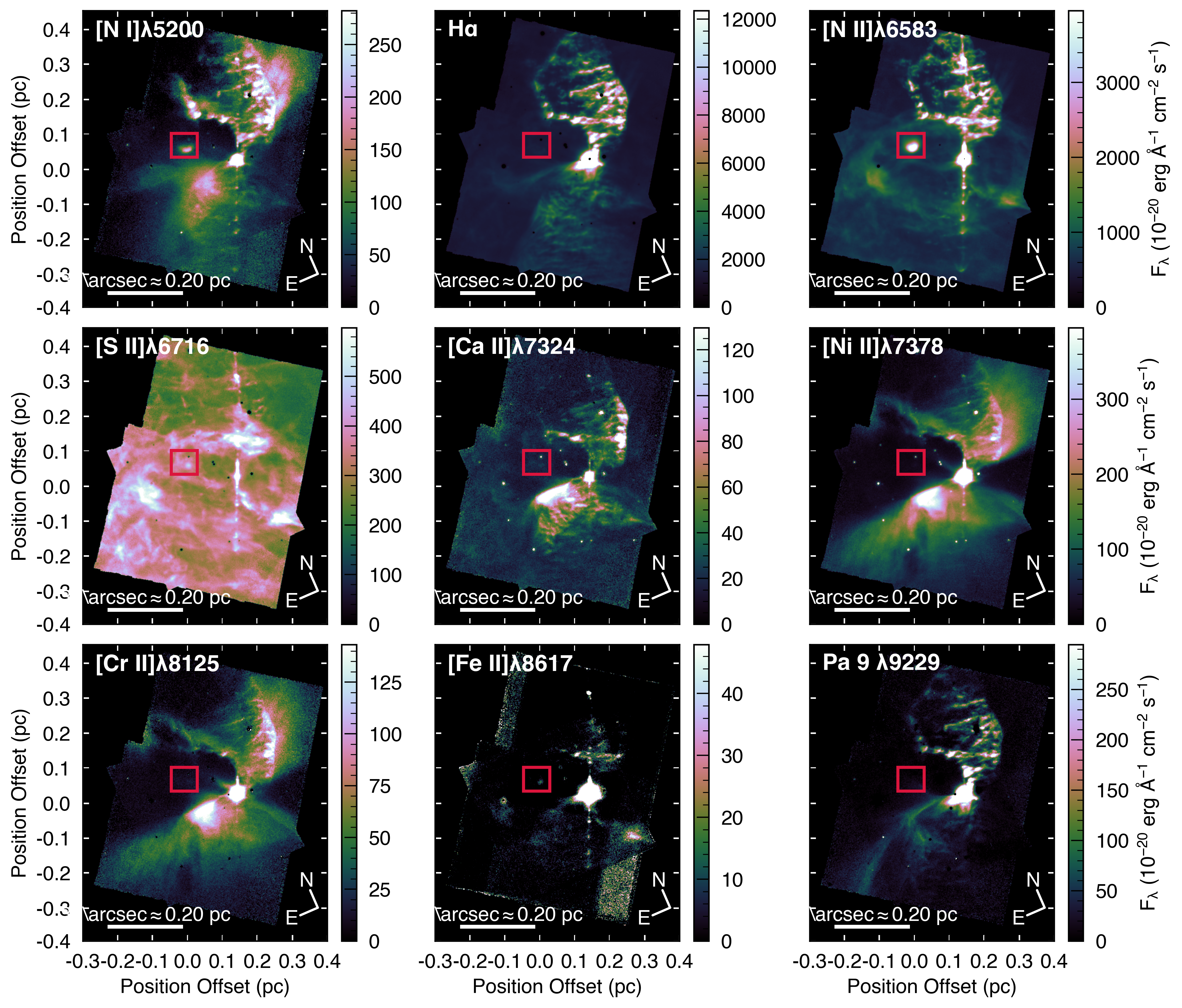}
\caption{Continuum-subtracted MUSE narrow-band images constructed by integrating each line over the velocity range $-150$~km\,s$^{-1}$ to $+200$~km\,s$^{-1}$. Top row: [N\,{\sc i}]$\lambda$5200, H$\alpha$, and [N\,{\sc ii}]$\lambda$6583 line emissions. Middle row: [S\,{\sc ii}]$\lambda$6716, [Ca\,{\sc ii}]$\lambda$7324, and [Ni\,{\sc ii}]$\lambda$7378. Bottom row [Cr\,{\sc ii}]$\lambda$8125, [Fe\,{\sc ii}]$\lambda$8617, and Pa 9 $\lambda$9229. The red square in each panel marks the position of Vela~Jr CCO nebula.}
\label{fig:muse_narrowBand}
\end{figure*}

\begin{figure*}[ht!]
\plotone{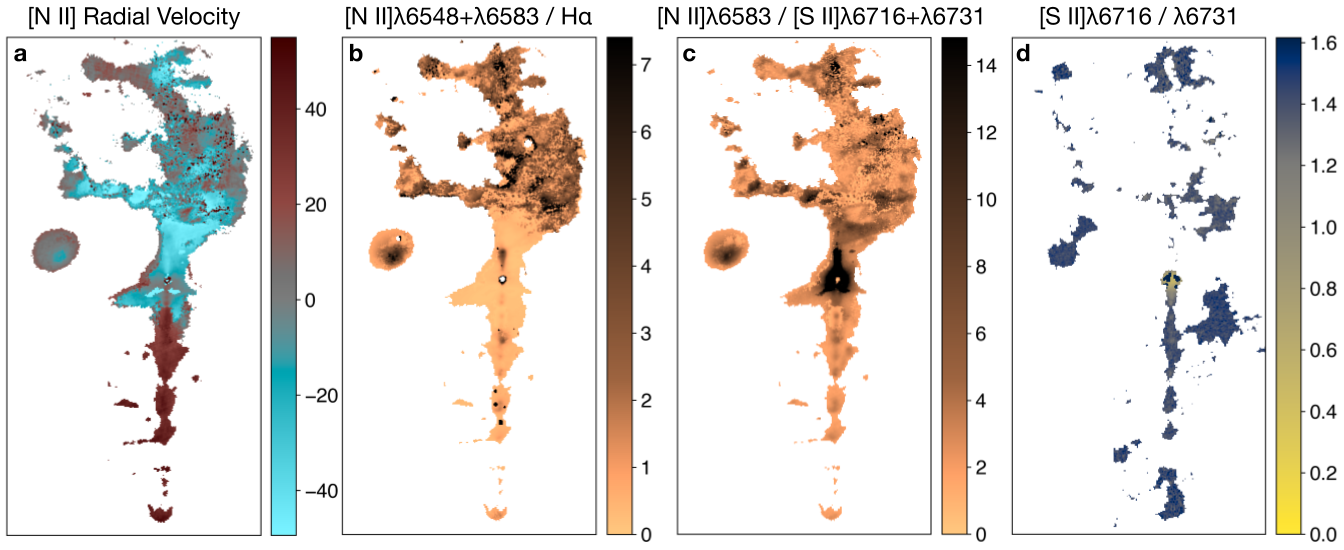}
\caption{(a) Velocity map of the [N\,{\sc ii}]$\lambda$6583 line, showing the kinematic structure of Ve~7--27 and Vela~Jr CCO nebula. (b) Spatial distribution of [N\,{\sc ii}]/H$\alpha$ ratio and (c) [N\,{\sc ii}]/[S\,{\sc ii}], both highlighting the strong nitrogen enrichment of both Ve~7--27 and the CCO nebula. (d) Spatial map of the canonical electron-density-sensitive [S\,{\sc ii}]$\lambda$6716/$\lambda$6731 ratio across the field. This ratio decreases with increasing electron density.}
\label{fig:ifu_maps}
\end{figure*}

The spatial distribution of individual emission lines is shown in Figure~\ref{fig:muse_narrowBand}. The jet knots exhibit the low-ionization emission ([N\,{\sc i}], [N\,{\sc ii}], [S\,{\sc ii}], and [Fe\,{\sc ii}]) that trace shock-excited gas arising from ambient material swept up and shocked by the jet, along with unusually strong Fe-peak emission ([Cr\,{\sc ii}] and [Ni\,{\sc ii}]) that points to an enriched environment. Along the outlfow, the shocked knots show exceptionally strong [N\,{\sc ii}] emission with notably no oxygen emission, whereas the central source of Ve~7--27 itself exhibits [N\,{\sc ii}]/H$\alpha<1$, indicating that the enrichment does not originate from the protostar or its natal cloud, but the environment that the jet is interacting with. The [N\,{\sc ii}]/H$\alpha$ and [N\,{\sc ii}]/[S\,{\sc ii}] maps (Figure~\ref{fig:ifu_maps}b and c) provide sensitive diagnostics of nitrogen enrichment and show consistently elevated values throughout both lobes of Ve~7--27. The northern cocoon is traced by Balmer and Paschen lines, [N\,{\sc i}], [N\,{\sc ii}], [S\,{\sc ii}], [Ca\,{\sc ii}], and [Fe\,{\sc ii}], while an extended hourglass-shaped nebula, shines particularly in [Ni\,{\sc ii}] and [Cr\,{\sc ii}], overlaps both lobes and outlines part of the western boundary of the northern cocoon.

A velocity map derived from multi-component fits to the [N\,{\sc ii}]$\lambda\lambda$6548,6583 lines (see Appendix~\ref{app:line_fitting}) in Figure~\ref{fig:ifu_maps}a shows a symmetric velocity gradient, with knot~D in the northern lobe reaching $\sim-50$~km\,s$^{-1}$ and the southernmost knot (knot A) reaching $\sim+50$~km\,s$^{-1}$. These velocities indicate a bipolar outflow, with one lobe approaching and the other receding, viewed at a moderate inclination to the line of sight. The spatial map of the density-sensitive [S\,{\sc ii}]$\lambda$6716/$\lambda$6731 ratio (Figure~\ref{fig:ifu_maps}d) shows that only a very compact region immediately surrounding the central source of Ve~7--27 exhibits high electron densities, with ratios $\lesssim0.5$ corresponding to $n_e\gtrsim10^4$~cm$^{-3}$ (for T$\approx10^4$~K). Outside this core, the ratios span 0.9--1.6, which corresponds to low-to-moderate densities of $n_e\lesssim500$~cm$^{-3}$ under typical nebular temperatures).

The observed collimated outflow, bipolar jet geometry, and shock-excited knots with bow shocks unambiguously point to Ve~7--27 as the driving source of a bipolar HH outflow.

\section{Revised Distance to Vela~Jr via Ve~7--27} \label{sec:revisedDistance}

Both the outflow of Ve~7--27 and the CCO nebula show emission characteristics that are not expected in normal interstellar or circumstellar environments. In classical HH objects, where jets propagate into molecular cloud material, shocks typically yield [N\,{\sc ii}]/H$\alpha$ $<$ 1 across a wide range of shock velocities and densities \citep[e.g.][]{Raga1996,Dopita2017}. However, the HH-shock of Ve~7--27 shows [N,{\sc ii}]/H$\alpha$ values reaching $\sim$7 (Figure~\ref{fig:ifu_maps}b), comparable to the extreme nitrogen enhancement observed in the CCO nebula. These values cannot be generated by shocks propagating into standard ISM compositions and indicate that the ambient gas is unusually nitrogen-rich.

To probe the nature of the nitrogen enrichment, we computed slow-shock models with \textsc{MAPPINGS v5.2.0} (see Appendix~\ref{app:mappings} for details). Figure~\ref{fig:mappings} compares MUSE measurements with the model predictions in the standard diagnostic plane of [S\,{\sc ii}]$\lambda$6761/$\lambda$6731 versus [N\,{\sc ii}]/[S\,{\sc ii}]$\lambda$6716+$\lambda$6731. Shock grids assuming standard ISM abundances \citep[Local Galactic Concordance; LGC;][]{Nicholls2017} fail to reproduce the observed [N\,{\sc ii}]/[S\,{\sc ii}] ratios for any combinations of shock velocities and pre-shock densities. In contrast, the observed values are well matched by models adopting nitrogen-enriched, WN-like abundances, identical to those used in \citet{Suherli2024} to model the photoionized CCO nebula. A complementary comparison using the [N\,{\sc ii}]/H$\alpha$ ratio is provided in Appendix~\ref{app:mappings} (Figure~\ref{fig:mappings_app}); although it is not a standard diagnostic diagram, it independently shows that the Ve~7--27 knots and the CCO nebula cannot be produced by shocks into gas of near ISM-composition.

\begin{figure}[ht!]
\plotone{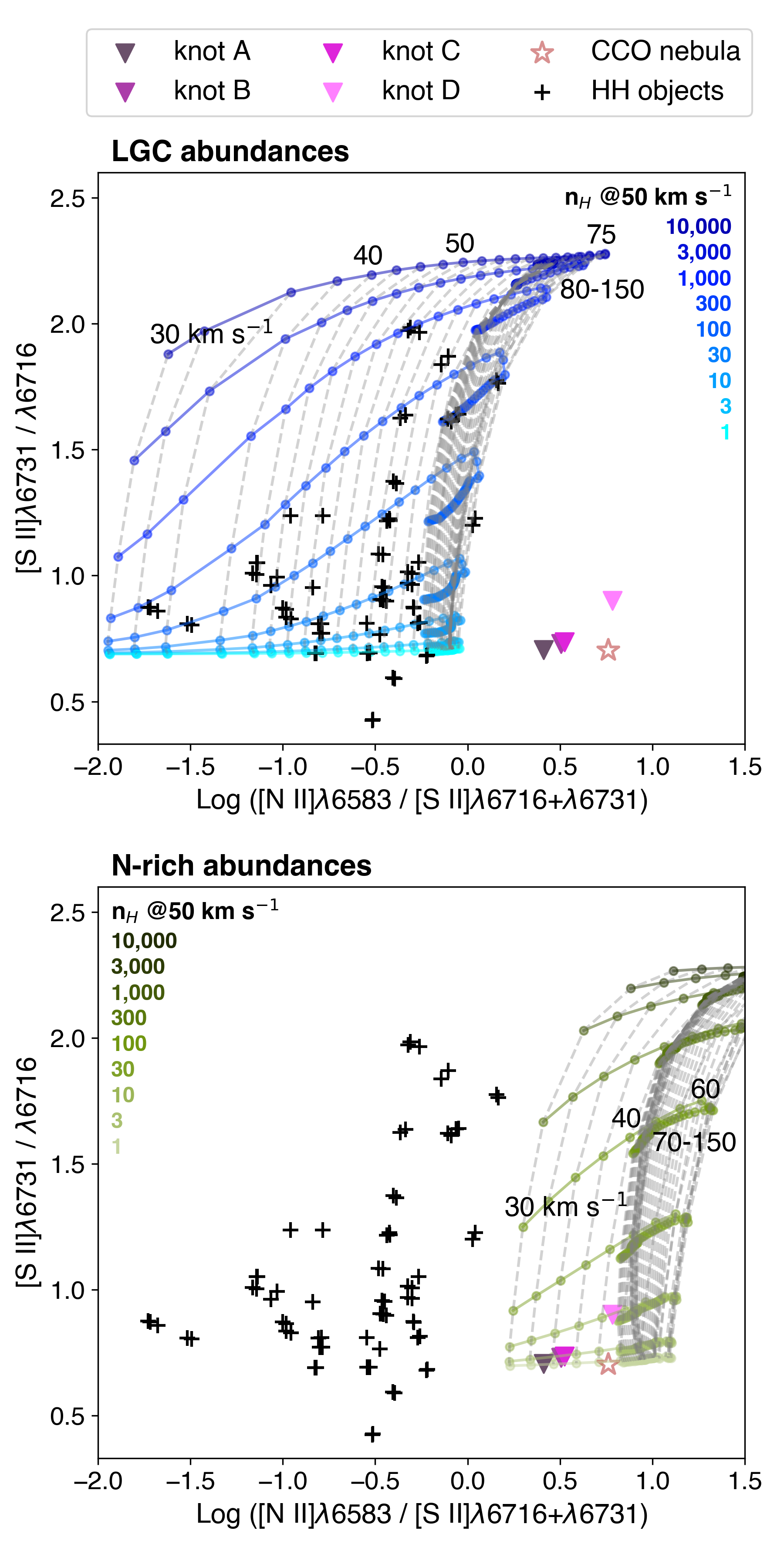}
\caption{Line-ratio diagnostic diagram of [S\,{\sc ii}]$\lambda$6716 / $\lambda$6731 versus [N\,{\sc ii}]$\lambda$6583 / [S\,{\sc ii}]$\lambda$6716+$\lambda$6731, comparing MUSE measurements with slow-shock model prediction, computed with Local Galactic Concordance (LGC) abundances in blue lines (top) and nitrogen-rich (WN-like) abundances from \citet{Suherli2024} in green lines (bottom). Each panel shows the predicted values for shock velocities of 30--150\,km\,s$^{-1}$ and pre-shock hydrogen densities spanning $n_{\rm H}=$ 1--10$^4$~cm$^{-3}$. Black `+' symbols mark the classical HH objects from \citep{Raga1996} and the inverted colored triangles represent HH knots A--D of Ve~7--27 (as labeled in Figure~\ref{fig:muse_annotated}). The hollow the red star denotes the CCO nebula; it is shown only for reference, as its emission is thought to arise from photoionization rather than shocks. }
\label{fig:mappings}
\end{figure}

The optical forbidden lines of Fe-peak elements ([Cr\,{\sc ii}], [Fe\,{\sc ii}], and [Ni\,{\sc ii}]) are rare or completely absent in typical HH objects, which usually show only infrared [Fe\,{\sc ii}] emission (e.g. at 1.64 $\mu$m). This is because young stellar environments, the protostars themselves and their surrounding molecular clouds, contain negligible gas-phase Fe-peak material, and protostars do not synthesize Fe-peak nuclei. Fe-peak elements are produced primarily in Type Ia supernovae or during the late evolutionary stages of massive stars, and they are released into the surrounding gas through supernova ejecta. Their strong presence in both Ve~7--27 and the CCO nebula (where [Fe\,{\sc ii}] emission is prominent) indicates that the gas being shocked by the Ve~7--27 jet is chemically pre-enriched, consistent with the ejecta of a massive progenitor rather than gas associated with a standard protostellar environment.

The bright northern cocoon of Ve~7--27 shows a clear geometric distortion along its eastern wall near the base of the outflow, precisely toward where the Vela~Jr CCO nebula is. Such deformation is not expected from the internal dynamics of an isolated HH flow, where in the absence of external forces or strong ambient gradients, the outflow maintain approximate symmetry about the flow axis \citep{Reipurth2001}. The same asymmetry appears in the Fe-peak hourglass shape traced by [Ni\,{\sc ii}] and [Cr\,{\sc ii}], whose northern lobe bends on the east side whereas the southern counterpart remains symmetric. This directional curvature seen across multiple emission lines indicates that an external influence, rather than intrinsic jet dynamics, is affecting the cocoon’s geometry.

Velocity channel maps constructed from H$\alpha$ (red), [N\,{\sc ii}]$\lambda$6583 (green), and [Ni\,{\sc ii}]$\lambda$7378 (blue) show that emission from Ve~7--27 and from the CCO nebula occupies the same spatial regions across a continuous sequence of velocity slices from $-150$~km\,s$^{-1}$ to $+200$~km\,s$^{-1}$ (Figure~\ref{fig:channelMaps}). The two objects do not appear as disjoint velocity structures, nor as unrelated components that overlap only in isolated bins, which is fully consistent with the [N\,{\sc ii}] velocity map (Figure~\ref{fig:ifu_maps}a), which shows that the HH knots and the CCO nebula span the same range of line-of-sight velocities. This spatial–kinematic continuity reinforces that both systems lie within the same physical volume of enriched gas, rather than being unrelated structures aligned by chance along the line of sight.

The combination of unusually strong nitrogen enrichment, shared morphology, and matched kinematics reveals more than a mere spatial coincidence between Ve~7--27 and the CCO nebula. The chemical signature of the WN-progenitor wind (enhanced nitrogen and depleted oxygen) imprinted on both objects links them to the same local environment, while the presence of Fe-peak forbidden lines indicates additional contribution from supernova ejecta. These results show that the Ve~7--27 jets are not plowing into ordinary ISM material, but into chemically processed material shed by the progenitor of Vela~Jr. The nitrogen-rich environment naturally accounts for the unusually high line ratios and highlights the unique nature of this system: an HH object interacting not with diffuse ISM, but with the enriched wind and ejecta of a dead massive star. Such an association has not been observed before, marking Ve~7--27 as an anomalously enriched HH object whose emission properties are directly shaped by the legacy of a massive supernova progenitor. This association anchors the CCO, and therefore Vela~Jr SNR, to the \textit{Gaia} parallax distance of Ve~7--27, yielding a revised distance of $1.41\pm0.14$~kpc.

\section{Discussion} \label{sec:discussion}

\subsection{Physical Properties of Vela~Jr at the Revised Distance}

Placing Vela~Jr at a distance of $1.41\pm0.14$~kpc through its association with Ve~7--27 (Figure~\ref{fig:distance_compilation}) resolves long-standing inconsistencies that have complicated the interpretation of the remnant's physical properties. With an angular radius of $0.95\pm0.02^{\circ}$ \citep{Camilloni2023}, Vela~Jr's physical radius becomes $23.3\pm2.3$~pc, which places Vela~Jr in line with other Galactic SNRs of comparable radio surface brightness \citep{Duncan2000}.

The only observational constraint on Vela~Jr's shock motion comes from proper motion studies of its northwestern rim, which report expansion rates of $0.84\pm0.23$~arcsec\,yr$^{-1}$ \citep{Katsuda2008} and $0.42\pm0.10$~arcsec\,yr$^{-1}$ \citep{Allen2015}. At the revised distance, these translate to shock velocities of $(5.6\pm1.5)\times10^3$~km\,s$^{-1}$ and $(2.8\pm0.7)\times10^3$~km\,s$^{-1}$, respectively. Assuming Sedov-Taylor evolution and $v=\frac{2}{5} R/t$, the inferred shock velocities imply ages of $\sim$1.6~kyr and $\sim$3.3~kyr. These values are in good agreement with the 2.4--5.1~kyr age range inferred by \citet{Allen2015} from hydrodynamical modeling and assuming expansion into a uniform ambient medium. 

Combining these ages with the Sedov-Taylor scaling relation, $R_s\simeq12.4~\mathrm{pc}~(E_{51}/n)^{0.2} t_4^{0.4}$, where $R_s$ is the remnant's radius, $E_{51}$ is the explosion energy in units of 10$^{51}$~ergs, $n$ the ambient density, and $t_4$ is the age is units of 10~kyr, we find $(E_{51}/n)\sim9\times10^2$ for $t\sim1.6$~kyr and $(E_{51}/n)\sim2\times10^2$ for $t\sim3.3$~kyr. For a canonical explosion energy of $E_{51}\sim1$, these values imply very low ambient densities of $n\sim10^{-3}-5\times10^{-3}$~cm$^{-3}$ -- consistent with expansion into a highly rarefied medium, such as a wind-blown cavity produced by a massive star progenitor. The inferred low ambient density explains the lack of strong thermal X-ray emission in Vela~Jr as well as the high inferred velocity from a few kyr-old remnant.

The revised distance also substantially increases the remnant’s inferred very-high-energy $\gamma$-ray luminosity, $L_\gamma(1-10~\mathrm{TeV})$, from $3\times10^{32}$~erg\,s$^{-1}$ \citep[][; that assumed a distance of 200~pc]{Aharonian2005} to $1.5\times10^{34}$~erg\,s$^{-1}$, placing Vela~Jr among the most luminous TeV-bright SNRs of comparable age. The corresponding total cosmic-ray proton energy above $1$~GeV rises to $W_p\approx5.1\times10^{48}$~erg, based on the scaling of \citealt{Fukui2024}, emphasizing Vela~Jr as an efficient cosmic-ray accelerator.

The implications extend further to the Vela~Jr CCO, which plays an important role in the study of the dense matter nuclear equation of state \citep[see, e.g.,][]{Potekhin2015}. At a distance of 750~pc, the inferred CCO's X-ray luminosity is $\sim$(1--2)$\times$10$^{32}$~erg~s$^{-1}$ \citep{Kargaltsev2002,Wu2021}, low enough at a presumed young age (2–5~kyr) to indicate rapid cooling via the direct Urca neutrino emission that is only predicted to occur in high mass neutron stars for certain equations of state \citep{Lattimer1991}. Now with almost twice as large a distance, the higher CCO luminosity is still lower than neutron stars with canonical mass that cool slowly (Ho et al., in prep.). The larger distance also increases the possibility that the CCO has a carbon surface, inferred from modeling of the CCO X-ray spectrum \citep{Potekhin2020,Wu2021,Alford2023}, although a carbon composition is still unlikely due to the CCO age \citep{Wijngaarden2019}.

\subsection{Linking Birth and Death}

Protostellar lifetimes \citep[$\sim$0.1--0.5~Myr;][]{Evans2009} and HH flow dynamical ages \citep[10$^{4}$--10$^{5}$~yr;][]{Reipurth2001} exceed the few-kyr age inferred for Vela~Jr, implying that Ve~7--27 must have formed well before the supernova explosion. Hydrodynamic simulations demonstrated that young stars and their protoplanetary disks can survive the blast wave of a nearby supernova, down to separations of a few tenths of a parsec, with significant stripping of the outer disks occurring only when the explosion is within $\lesssim0.1$~pc \citep[][and references therein]{Ouellette2010}. At the revised distance, the projected 22\arcsec\ separation between Ve~7--27 and the Vela~Jr CCO corresponds to $\sim0.15$~pc, which is in a range where disks are expected to survive, bearing in mind that the true distance is almost certainly greater than the projected value. Thus, a system such as Ve~7--27 could plausibly survive the Vela~Jr supernova explosion. 

The progenitor's WN wind created a nitrogen-rich, oxygen-poor environment, and the subsequent core-collapse explosion enriched the same region with Fe-peak material. These abundance patterns observed in both Ve~7--27 and in the CCO nebula (Figures~\ref{fig:muse_narrowBand} and~\ref{fig:ifu_maps}b,c) indicate that Ve~7--27 system resides within the chemically-enriched medium produced by the massive progenitor of Vela~Jr during its final evolutionary stages. 

The relatively small Doppler velocities observed in both the CCO nebula and the Ve~7--27 jet knots (Figure~\ref{fig:ifu_maps}a) further support this picture. For HH objects to produce observable emission, their jets must collide with ambient material dense enough to generate radiative shocks. Such conditions arise naturally in the late WN phase, when the winds become increasingly clumpy and their velocities decline from thousands of km\,s$^{-1}$ to only a few tens of km\,s$^{-1}$ near core collapse (\citealt{Crowther2007}; see evolutionary models by \citealt{Roy2021}). This slow, nitrogen-rich material provides an ideal environment for producing the modest jet velocities observed and for sustaining the strong shock-excited emission from the knots and cocoon.

\section{Conclusions} \label{sec:conclusions}

Accurately determining distances to Galactic supernova remnants remains a fundamental challenge, particularly for systems dominated by non-thermal emission. In Vela~Jr, the chemical, spatial, and kinematic link between the CCO nebula and Ve~7--27 provides the first evidence of a YSO co-evolving within supernova-processed material. The Gaia parallax of Ve~7--27 therefore provides a direct geometric distance to the Vela~Jr SNR, resolving the long-standing distance ambiguity for this enigmatic remnant and enabling a substantial revision of its physical properties. 

The key results of this study are as follows:
\begin{itemize}
    \item Ve~7--27 is a Young Stellar Object driving a bipolar Herbig-Haro outflow, located within the central $\sim$2\arcmin \,of the Vela~Jr Supernova Remnant.
    \item Ve~7--27 and the Vela~Jr CCO nebula share similar abundance signatures of unusually strong nitrogen enrichment and Fe-peak enhancement, confirming that they occupy the same chemically processed environment rather than a chance alignment.
    \item This association places Vela~Jr at the distance of $1.41\pm0.14$~kpc, resolving decades of inconsistent distance estimates.
    \item At this distance, Vela~Jr has a physical radius of $23.3\pm2.3$~pc and an age of $\sim$~1.6--3.3~kyr, based on the proper motion measurements of the northwestern rim. Its large radius, high shock velocities, and very low inferred ambient density together indicate expansion into a wind-blown cavity carved by a massive progenitor. This naturally explains the remnant's lack of strong thermal X-ray emission and the high blast wave velocity inferred for a 2--3 kyr old remnant dominated by non-thermal X-ray emission -- further highlighting Vela Jr's importance as a bright gamma-ray source and efficient particle accelerator. 
\end{itemize}

These findings present a coherent picture of Vela~Jr as a rapidly expanding SNR evolving within a low-density cavity, whose central region is interacting with an unusually enriched HH outflow, providing a rare, direct view of how an SNR can influence nearby young stellar activity. Further optical and infrared spectroscopy, particularly with \textit{James Webb Space Telescope} (JWST), will refine the abundance measurements of the Fe-peak elements and further constrain the excitation mechanisms in this unique environment. \textit{Imaging X-ray Polarimetry Explorer} (IXPE) observations will shed light on the magnetic field geometry in the fast moving shocks and the processes for efficient particle acceleration in this non-thermal dominated SNR. Upcoming high-energy facilities such as the \textit{Advanced X-ray Imaging Satellite} \citep[AXIS;][]{2023SPIE12678E..1ER,2023arXiv231107673S} and the \textit{Southern Wide-Field Gamma-ray Observatory} \citep[SWGO;][]{2019arXiv190208429A} will provide powerful new constraints on shock physics, particle acceleration, and the complex environments that govern the evolution of Vela~Jr and similar SNRs.

\begin{acknowledgments}
We thank G. Ferrand for insightful discussions on visualization and SNR evolutionary scenarios.
We also thank the anonymous referee for their careful review and valuable suggestions.
This research made use of NASA's ADS and SNRcat, the high-energy catalog of supernova remnants (\url{http://snrcat.physics.umanitoba.ca}). 
J.S. and S.S.H. acknowledge support from the Natural Sciences and Engineering Research Council of Canada (NSERC) through the Discovery Grants and the Canada Research Chairs programs.
A.J.R. acknowledges funding support from the Australian Research Council, Future Fellowship award FT170100243.
C-J.L. is supported by the NSTC grant 114-2112-M-004-001-MY2 from the National Science and Technology Council of Taiwan.
R.M.C. acknowledges funding from the Australian Research Council under grant DP230101055.
\end{acknowledgments}

\begin{contribution}

J.S. led the project, performed the data analysis and modeling, and wrote most of the text. I.R.S. and S.S.H. provided scientific guidance and contributed to the interpretation of the results and editing of the manuscript. F.P.A.V. led the MUSE observing proposal and acquired the data presented in this study. All authors contributed to the discussion of the results and reviewed the manuscript.


\end{contribution}

%
\facility{VLT:Yepun (MUSE)}

\software{
            \textsc{Scipy} \citep{2020SciPy-NMeth},
            \textsc{Astropy} \citep{2013A&A...558A..33A,2018AJ....156..123A,2022ApJ...935..167A},
            \textsc{brutifus} \citep{brutifus} (a Python module to process datacubes
                from integral field spectrographs, that relies on \textsc{statsmodel} \citep{seabold2010}, \textsc{matplotlib}, \textsc{astropy}, and \textsc{photutils}, an affiliated package of \textsc{astropy} for photometry), 
            \textsc{Matplotlib} \citep{Hunter:2007} (including the \textsc{cubehelix} colormap \citep{cubehelix}), and
            \textsc{CosmosCanvas} \citep{CosmosCanvas}.
          }


\appendix

\section{Published Distance Estimates for Vela~Jr} \label{app:distance_review}

Early distance estimates for Vela~Jr relied on X-ray surface brightness \citep{Aschenbach1998} and a detection of excess $^{44}$Ti $\gamma$-ray emission \citep{Iyudin1998,Aschenbach1999}, which implied an exceptionally young and nearby remnant at 200--500~pc. Such proximity would require implausibly small physical dimensions and was inconsistent with the remnant’s unusually low radio surface brightness, placing it well below the empirical $\Sigma$--$D$ relation for Galactic SNRs \citep{Duncan2000}. Later work showed that the reported $^{44}$Ti feature was weak and ambiguous, making it unreliable as a distance indicator \citep{Renaud2006,Weinberger2020}. More recent constraints come on X-ray absorption modeling or shock expansion measurements \citep{Slane2001,Katsuda2008,Acero2013,Allen2015} favor a larger distance but are inherently model-dependent and remain sensitive to assumptions about foreground absorption, ambient density, and explosion energy. In addition, Vela~Jr’s position in a crowded and complex region complicates the separation of its emission components, particularly in radio and X-ray analyses. As a result, most existing distance estimates rely on indirect diagnostics tied to specific model assumptions and often focus on the bright northwestern rim, which may not represent the remnant as a whole.

Attempts to use the CCO as a distance anchor for Vela~Jr have yielded no definitive result. \textit{Chandra} proper motion measurements remain inconclusive due to the scarcity of suitable background reference sources, and a two-epoch comparison shows no significant displacement, implying a proper motion statistically consistent with zero \citep{Mignani2019,Camilloni2023}. This reinforces the view that the CCO remains near its birth site at the geometric center of the remnant. An indirect estimate of 2.4$\pm$0.4~kpc \citep{Reynoso2006} based on X-ray absorption modeling depends strongly on assumptions about Galactic gas distribution and therefore cannot robustly fix the distance.

\section{MUSE Data} \label{app:muse_data_desc}

A detailed description of the MUSE Wide Field Mode (WFM) program (Program ID 0104.D-0092(B); P.I.: F.P.A. Vogt), data reduction, calibrations, and post-processing sequences, is provided in Section 2 of \cite{Suherli2024}. While the earlier publication focused on a 10\arcsec$\times$10\arcsec\,region centered on the CCO, in this work we use the full field of the final datacube. The complete mosaic spans $607\times623$~spaxels, corresponding to an area of approximately 121.4\arcsec$\times$124.6\arcsec\,on the sky, at the MUSE spatial sampling of 0.2\arcsec\,spaxel$^{-1}$.

\section{Emission-line fitting and velocity channel maps} \label{app:line_fitting}

To derive the kinematic and line-ratio maps presented in this work, we performed a spaxel-by-spaxel multi-Gaussian fit to the prominent optical emission lines within the CCO Nebula and Ve~7--27 HH outflow region of the MUSE datacube. The analysis was carried out using a dedicated \textsc{Python} workflow that applies single- and two-component Gaussian fits, optimized with a non-linear least squares fitter (\textsc{scipy}'s \textsc{curve\_fit} function), to H$\alpha$, [N\,{\sc ii}]$\lambda\lambda$6548,6583 doublet (with the theoretical 1:3 flux ratio enforced), and [S\,{\sc ii}]$\lambda\lambda$6716,6731 doublet. Two-component fits were attempted only when the single-component model failed to provide an adequate fit of the spectrum, otherwise, spaxels defaulted to a single-component solution. Instrumental broadening at each wavelength was determined from the MUSE line-spread function provided by \textsc{brutifus} \citep{brutifus}, and the instrumental $\sigma$ was adopted as the lower bound for all the fitted line widths.

In addition to the spaxel-by-spaxel analysis, we extracted integrated spectra for knots A--D and for the CCO nebula, as indicated in Figure~\ref{fig:muse_annotated}. Each spectrum was obtained by summing over a $7\times4$~spaxel region centered on the peak-brightness spaxel of each feature. The integrated spectra were fitted using simultaneous two-component Gaussian models for the H$\alpha$, [N\,{\sc ii}]$\lambda\lambda$6548,6583, and [S\,{\sc ii}]$\lambda\lambda$6716,6731 emission lines. Line fitting was carried out using a custom \textsc{Python} code that implements the \textsc{Scipy}'s \textsc{curve\_fit} function, which minimizes the residuals between the observed line profiles and the model. Simultaneously fitting all five lines provides robust measurements of the line fluxes and Doppler velocities of both Gaussian components. From the fitted fluxes, we computed the [N\,{\sc ii}]/H$\alpha$ and [N\,{\sc ii}]/[S\,{\sc ii}] ratios, which serves as an indicator of nitrogen enrichment, as well as the [S\,{\sc ii}]$\lambda$6731/$\lambda$6716 ratio.

\begin{figure*}[ht!]
\plotone{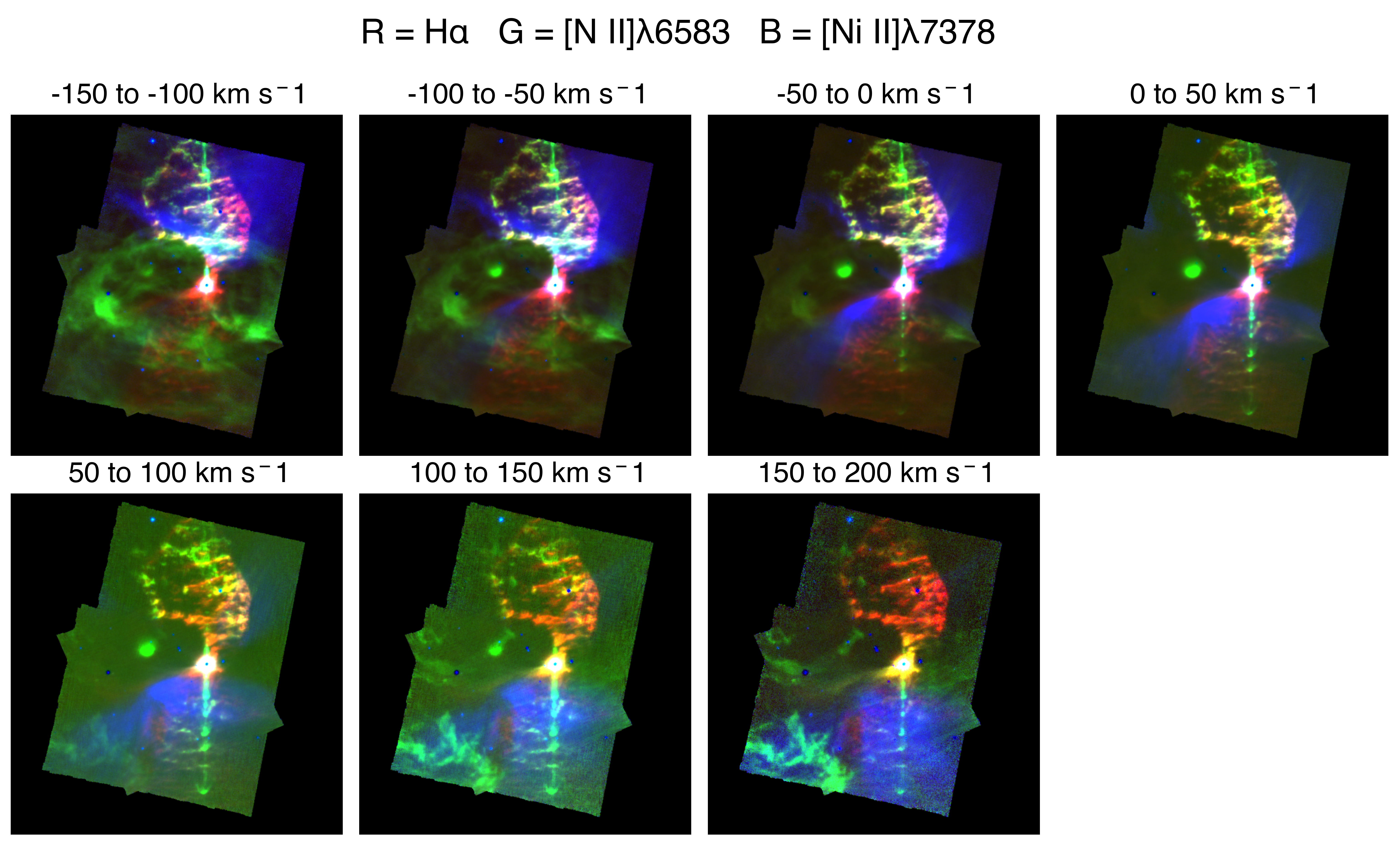}
\caption{Channel maps of H$\alpha$ (red), [N~{\sc ii}]6583 (green), and [Ni~{\sc ii}]7378 (blue) emission from the continuum-subtracted MUSE datacube, from $-150$ to $+200~\mathrm{km,s^{-1}}$. Each panel shows the integrated intensity within a fixed velocity slice indicated in the label. The red square marks the CCO nebula.}
\label{fig:channelMaps}
\end{figure*}

Channel maps are two-dimensional images that isolate emission within specific velocity intervals along the line of sight, allowing for the spatial distribution of gas at different radial velocities to be visualized. We constructed channel maps of H$\alpha$ (red), [N\,{\sc ii}]$\lambda$6583 (green), and [Ni~{\sc ii}]$\lambda$7378 (blue) emission (Figure~\ref{fig:channelMaps}) from the continuum-subtracted MUSE datacube.

\section{Shock Modeling with MAPPINGS} \label{app:mappings}

We performed slow-shock modeling using \textsc{MAPPINGS v5.2.0} \citep{mappings}, considering both standard ISM (Local Galactic Concordance, LGC) abundances \citep{Nicholls2017} and nitrogen-rich abundances of WN star evolution \citep[as adopted in][]{Suherli2024}, over a velocity range of 30 to 250~km\,s$^{-1}$ in increments of 2.5~km\,s$^{-1}$, using a fixed shock ram pressure to eliminate the velocity dependence on surface brightness. The ram pressure is normalized relative to the pre-shock hydrogen number density at a shock velocity at 50~km\,s$^{-1}$. Constant-pressure conditions were modeled over the entire velocity range for pre-shock hydrogen densities ($n_{\rm H}$) of 1, 3, 10, 30, 100, 300, 1000, 3000, and 10,000~cm$^{-3}$.

\begin{figure}[ht!]
\centering
\includegraphics[width=0.4\linewidth]{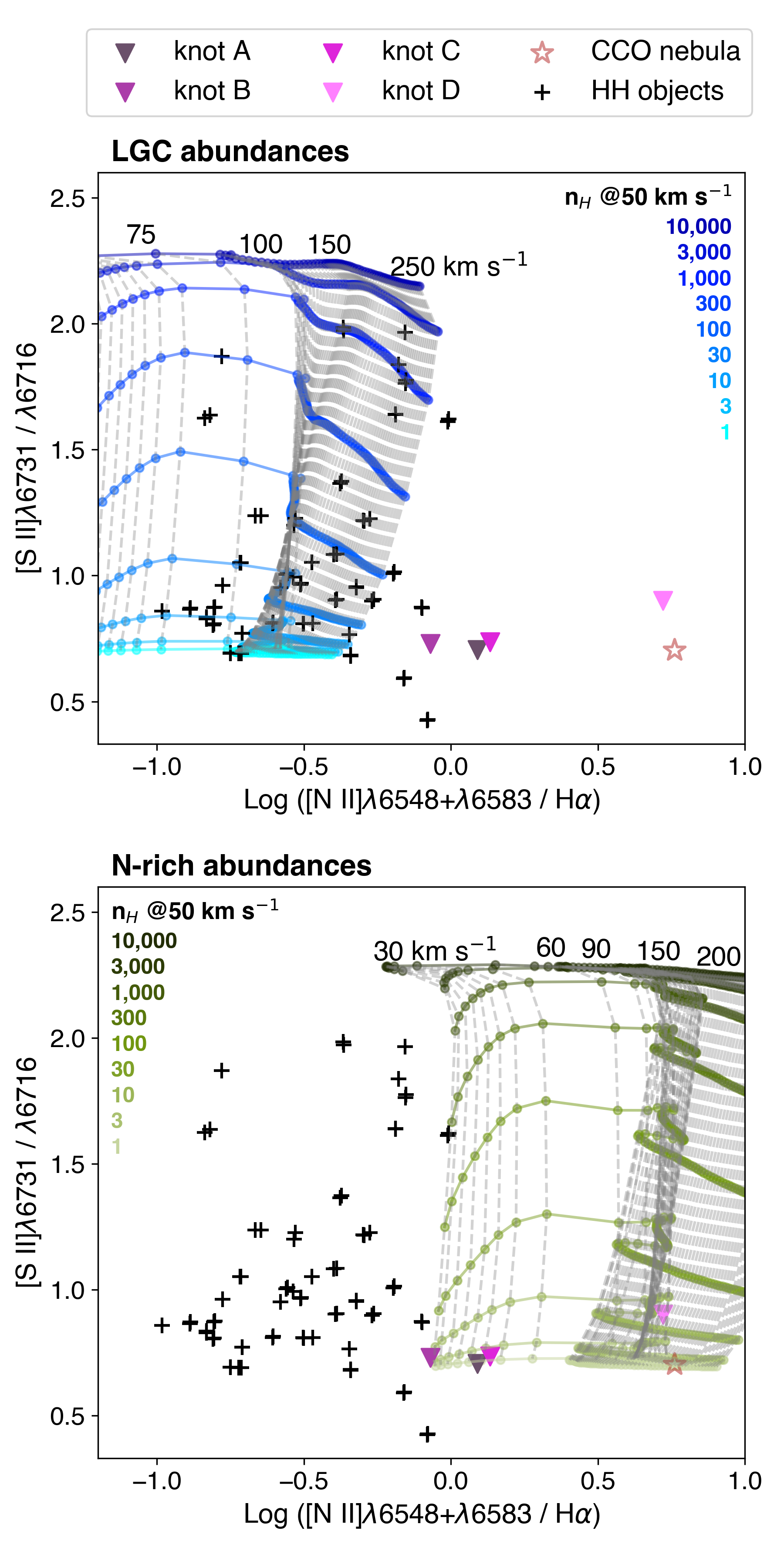}
\caption{Line-ratio diagnostic diagram of [S\,{\sc ii}]$\lambda$6716 / $\lambda$6731 versus [N\,{\sc ii}]$\lambda$6548+$\lambda$6583 / H$\alpha$, comparing MUSE measurements with slow-shock model prediction, computed with Local Galactic Concordance (LGC) abundances in blue lines (top) and nitrogen-rich (WN-like) abundances from \citet{Suherli2024} in green lines (bottom). Each panel shows the predicted values for shock velocities of 30-250~km\,s$^{-1}$ and pre-shock hydrogen densities spanning $n_{\rm H}=1-10^4$~cm$^{-3}$. Black `+' symbols mark the classical HH objects from \citep{Raga1996} and the inverted colored triangles represent HH knots A--D of Ve~7--27 (as labeled in Figure~\ref{fig:muse_annotated}). The hollow the red star denotes the CCO nebula; it is shown only for reference, as its emission is thought to arise from photoionization rather than shocks. }
\label{fig:mappings_app}
\end{figure}

\section{Total energy (in cosmic-ray protons) approximation} \label{app:Wp}

In hadronic models, where $\gamma$-rays arise from collisions between shock-accelerated protons and ambient gas, the $\gamma$-ray luminosity $L_\gamma$ is governed by the total energy in cosmic-ray protons ($W_p$) and the proton–proton collision timescale ($t_{pp}$), such that $L_\gamma \propto W_p/t_{pp}$. Here, $t_{pp}$ is inversely proportional to the ambient hydrogen density $n_H$. The ambient density can be approximated by $n_H \approx N_H/\Delta D$, where $N_H$ is the hydrogen column density and $\Delta D$ is the physical path length through the emitting region. To the extent that we can ignore other gas along the line of sight, the $N_H$ is a measured quantity, and given we can reasonable adopt that $\Delta D \propto d$ at fixed angular size, 
therefore the collision timescale scales with distance as $t_{pp} \propto 1/n_H \propto \Delta D/N_H \propto d$.

Given that $L_\gamma \propto d^2 F_\gamma$, we can derive the scaling relation for $W_p$ as $L_\gamma \propto W_p/d \Longrightarrow W_p \propto d^3$. With this cubic dependence on distance, the revised distance of 1.41~kpc implies that Vela~Jr’s total CR proton energy is a factor of $(1.41/0.75)^3 \approx 6.6$ higher than previously estimated. This raises the inferred $W_p$ from $7.7 \times 10^{47}$~erg (for $>$1 GeV cosmic ray protons), assuming $d=0.75$~kpc and $n_H = 90$~cm$^{-3}$ \citep{Fukui2024}, to roughly $5.1 \times 10^{48}$~erg, exceeding the value estimated for RX~J1713.7--3946 by at least a factor of five. This suggests that Vela~Jr is one of the most powerful Galactic cosmic-ray accelerators among known Galactic SNRs


\bibliography{VelaJrDistance_ApJL}{}
\bibliographystyle{aasjournal}



\end{document}